\documentclass[runningheads]{llncs}
\usepackage{amsfonts}
\usepackage{graphicx}
\usepackage{amsfonts}
\usepackage{graphicx}
\usepackage{epsfig}
\usepackage{graphicx}
\usepackage{amsmath}
\usepackage{amssymb}
\usepackage{mathtools}
\usepackage{resizegather}
\usepackage{marvosym}
\usepackage{dsfont}
\usepackage{enumitem}
\usepackage{caption}
\usepackage{multirow}
\RequirePackage[utf8]{inputenc}

\usepackage[colorlinks=true, allcolors=blue]{hyperref}

\usepackage[ruled, vlined]{algorithm2e}

\usepackage{lipsum}
\usepackage{array}
\usepackage{amsmath}

\usepackage{enumitem}
 
\begin{document}
%
%\title{COST: Unsupervised Domain Adaptation with Complementary Self-Training for Segmentation}
\title{Tagged-MRI Sequence to Audio Synthesis via Self Residual Attention Guided Heterogeneous Translator}

%\title{Self Residual Attention Guided Heterogeneous Translator for tagged-MRI Sequence to Audio Synthesis}
%

\titlerunning{Tagged-MRI2Audio with Attention Guided Heterogeneous Translator}

\author{Xiaofeng Liu\inst{1} \and Fangxu Xing\inst{1} \and Jerry L. Prince\inst{2} \and Jiachen Zhuo\inst{3} \and Maureen Stone\inst{3} \and  Georges El Fakhri\inst{1} \and Jonghye Woo\inst{1}}
 %\\ Anonymous Submission 
%\author{Xiaofeng Liu\inst{1} \and Fangxu Xing\inst{1} \and Chao Yang\inst{2} \and Georges El Fakhri\inst{1}  \and~~~ Jonghye Woo\inst{1}}

%index{Xiaofeng, Liu}
%index{Fangxu, Xing}
%index{Jerry L., Prince}
%index{Jiachen, Zhuo}
%index{Maureen, Stone}
%index{Georges, El Fakhri}
%index{Jonghye, Woo}

\institute{Massachusetts General Hospital and Harvard Medical School, Boston, MA, USA\and
Johns Hopkins University, Baltimore, MD, USA\and
University of Maryland, Baltimore, MD, USA 
}

\authorrunning{X. Liu et al.}

\maketitle              % typeset the header of the contribution

\begin{abstract}

Understanding the underlying relationship between tongue and oropharyngeal muscle deformation seen in tagged-MRI and intelligible speech plays an important role in advancing speech motor control theories and treatment of speech related-disorders. Because of their heterogeneous representations, however, direct mapping between the two modalities---i.e., two-dimensional (mid-sagittal slice) plus time tagged-MRI sequence and its corresponding one-dimensional waveform---is not straightforward. Instead, we resort to two-dimensional spectrograms as an intermediate representation, which contains both pitch and resonance, from which to develop an end-to-end deep learning framework to translate from a sequence of tagged-MRI to its corresponding audio waveform with limited dataset size.~Our framework is based on a novel fully convolutional asymmetry translator with guidance of a self residual attention strategy to specifically exploit the moving muscular structures during speech.~In addition, we leverage a pairwise correlation of the samples with the same utterances with a latent space representation disentanglement strategy.~Furthermore, we incorporate an adversarial training approach with generative adversarial networks to offer improved realism on our generated spectrograms.~Our experimental results, carried out with a total of 63 tagged-MRI sequences alongside speech acoustics, showed that our framework enabled the generation of clear audio waveforms from a sequence of tagged-MRI, surpassing competing methods. Thus, our framework provides the great potential to help better understand the relationship between the two modalities.

\end{abstract}

\section{Introduction} 
%The multimodal data,  that describe the same event can provide complementary information to under

% Xiaofeng, add any part from these sentences that you then think is helpful.
%%%%%%%%%%%%%%%%%%%%%%%%%%%%%
To facilitate our understanding of speech motor control in healthy and disease populations, associating dynamic imaging data with speech audio waveforms is an essential step in identifying the underlying relationship between tongue and oropharyngeal muscle deformation and its corresponding acoustic information \cite{liu2022cmri2spec,liu2022tagged,yu2021reconstructing}. Naturally, audio data recorded during scanning sessions need to be strictly paired and temporally synced with the dynamic tagged-MRI data to maintain their underlying relationship. However, when examining such paired data, we often face the following difficulties: 1) recorded audio data may be lost or non-existent from previously established tagged-MRI protocols; 2) audio waveforms may be heavily corrupted from scanner noise or poorly controlled noise-reduction protocols; and 3) time-stamps between tagged-MRI and its audio recording do not match and cannot be easily reconstructed without heavy manual intervention. All of these situations prevent pairing of the two data sources and cause missing, partial, or incomplete audio data. Therefore, recovering audio from imaging data itself becomes a necessary topic to be explored.
 
Heterogeneous data representations between the low-frame rate image sequence (i.e., two-dimensional (mid-sagittal slice) plus time (2D+t) tagged-MRI sequence) and high-frequency one-dimensional (1D) audio waveform make their translation a challenging task \cite{chung2016lip,liu2021generative,liu2022structure,xing2022brain,liu2020symmetric}. Besides, cross-modality speech models often lose pitch information \cite{akbari2018lip2audspec,ephrat2017vid2speech}. In contrast, a two-dimensional (2D) mel spectrogram converted from its 1D audio waveform represents the energy distribution of audio signals over the frequency domain along the time axis, which contains both pitch and resonance information of audio signals \cite{akbari2018lip2audspec,ephrat2017vid2speech,he2020image2audio,he2020classification}. Then, mel spectrograms can be converted back into audio waveforms \cite{griffin1984signal}. Accordingly, in this work, we opt to use the 2D mel spectrograms as an intermediate means to bridge the gap between tagged-MRI sequences and audio waveforms. This strategy has been applied to cine-MRI to spectrogram synthesis \cite{liu2022cmri2spec}.

As a related development in lip reading, early works have focused on using convolutional neural networks (CNN) and linear predictive coding analysis or line spectrum pairs decomposition~\cite{ephrat2017vid2speech,michelsanti2021overview} to achieve the goal. These methods, however, failed to maintain the frequency and periodicity of waveforms. To alleviate this, Lip2AudSpec~\cite{akbari2018lip2audspec} adopted a recurrent neural network (RNN) for the temporal modeling of CNN features, followed by applying fully connected layers to carry out the synthesis of spectrograms from 2D+t videos. However, it is challenging to train RNNs in general~\cite{pascanu2013difficulty,xie2021deep}, thereby likely yielding a suboptimal solution, especially when the network is trained on a limited number of datasets~\cite{wang2021automated}. In addition, the fully connected layers employed in Lip2AudSpec cannot capture spatial and temporal correlations~\cite{liu2020auto3d}, and demand massive to-be-trained parameters~\cite{goodfellow2017deep}. Furthermore, image sequences for audio analysis have a large amount of regions that are redundant or irrelevant for voice generation.

To sidestep these issues, we propose a self residual attention-guided heterogeneous translator to achieve an efficient tagged-MRI sequence to spectrogram synthesis. Specifically, we use an end-to-end asymmetric fully convolutional encoder-decoder translator as our backbone. Without the use of RNN and fully connected layers, our network has 10$\times$ fewer network parameters than Lip2AudSpec~\cite{akbari2018lip2audspec}. To explicitly exploit the movements of the tongue and vocal cords, the residual of adjacent frames are processed via a self-learned attention network to render the attention masks to guide the information extraction process. With our framework, we are able to largely eliminate the redundant regions in the tagged-MRI sequence, which allows us to bypass an additional delineation effort. Furthermore, an additional optimization constraint in latent space is imposed following a prior knowledge that part of the feature representation of the same utterance (e.g., ``ageese” or ``asouk” in this work) are similar to each other. To facilitate the disentanglement of utterance-specific and subject-specific factors in our fully convolutional network, we adopt a tensor slice operation with an information bottleneck to achieve the separation. For pairs with the same utterance, we explicitly enforce their utterance content part as close as possible using the Kullback-Leibler (KL) divergence.~Then, the decoder takes into account both the utterance content and the style of the articulation for the spectrogram synthesis. In addition, a generative adversarial network (GAN) loss \cite{salimans2016improved} is added on to yield improved realism on the synthesized spectrograms. 

The main contributions of this work can be summarized as follows:

\noindent$\bullet$ To our knowledge, this is the first attempt at translating tagged-MRI sequences to audio waveforms.%, which can potentially benefit clinicians and researchers in better understanding the correlation between the two different modalities and in potentially improving treatment strategies for patients with speech-related disorders.  

\noindent$\bullet$ We propose a novel self residual attention guided heterogeneous translator to achieve efficient tagged-MRI-to-spectrogram synthesis. 

\noindent$\bullet$ The utterance and subject factors disentanglement and adversarial training are further explored to improve synthesis performance.

Both quantitative and qualitative evaluation results using a total of 63 participants demonstrate superior performance in synthesis over comparison methods.

%The generative adversarial network (GAN) loss is simply added on to further boost the performance. In addition to a vanilla encoder-decoder, we further propose to exploit the similarity of the same word content in a latent space and to enforce an additional constraint.

%***Why MRI than lip video?:
%For example, it is nearly impossible to visually distinguish `s' from `z' as they are both pronounced with tongue near the gums and differ by the presence of voicing or vibration of the vocal cords in one (in the case of `z') as opposed to the other (here, `s').

\begin{figure}[t]
\begin{center}
\includegraphics[width=1\linewidth]{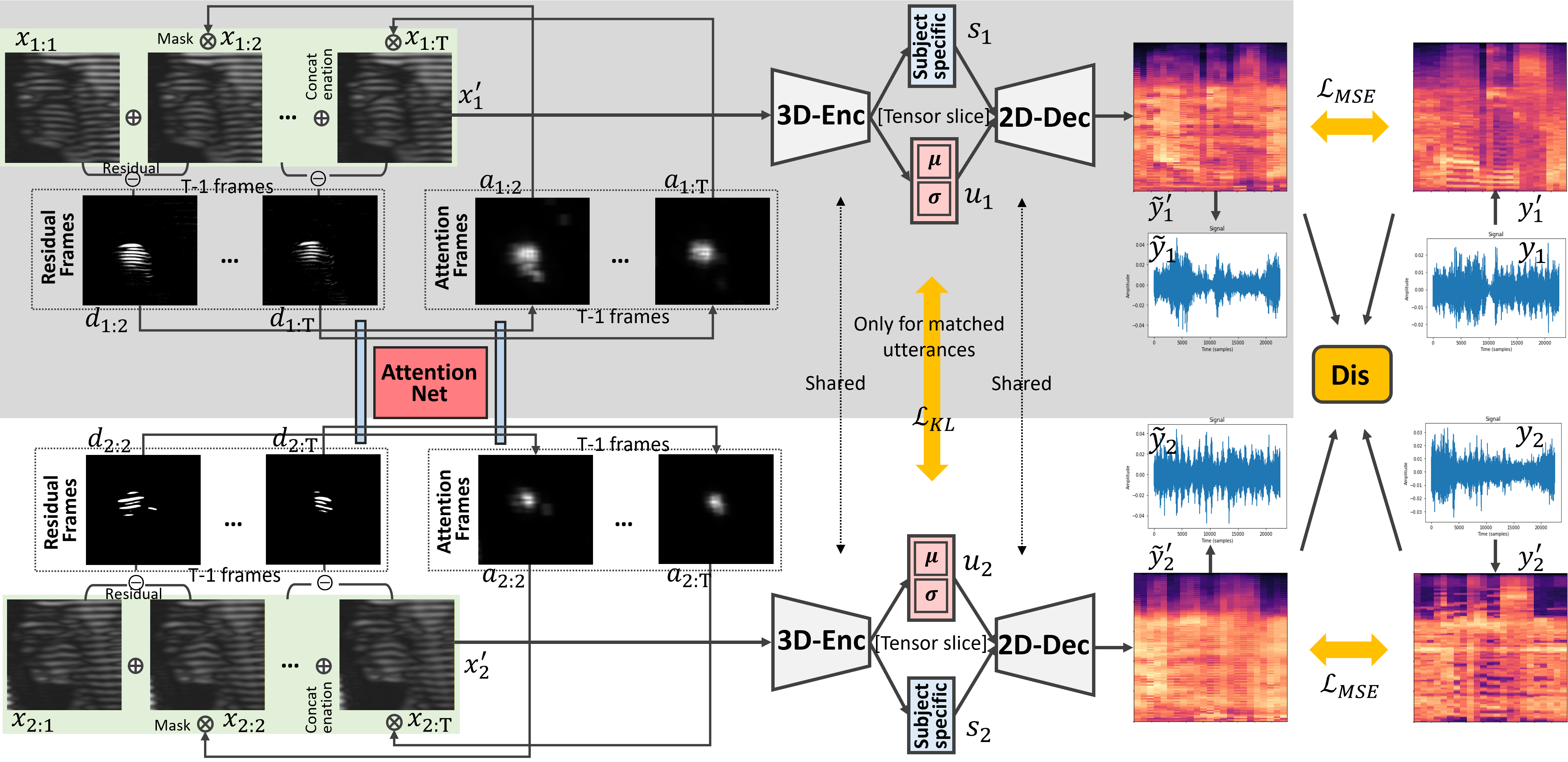}    
\end{center} 
\caption{Illustration of our self residual attention guided heterogeneous translator with latent space disentanglement and adversarial training. Only the gray masked modules are used at the testing stage.} 
\label{fig:illus}
\end{figure}

\section{Methodology}

Given a set of training pairs $\{x_i, y_i\}_{i=1}^N$ with a sequence of tagged-MRI $x_i\in\mathbb{R}^{H_x\times W_x\times T_x}$ and its corresponding quasi-synchronous audio spectrogram $y_i\in\mathbb{R}^{1\times L}$, we target to learn a parameterized heterogeneous translator $\mathcal{F}:x_i\rightarrow \tilde{y}'_i$ to generate a spectrogram $\tilde{y}'_i\in\mathbb{R}^{H_s\times W_s}$ from $x_i$, with the aim to resemble the spectrogram ${y}'_i$ of $y_i$. We denote the height, width and frames of $x_i$ as $H_x$, $W_x$, and $T_x$, respectively. In addition, $L$ is the length of the waveform $y_i$, and $H_s\times W_s$ indicates the size of the spectrogram. In this work, all of the audio waveforms are converted into mel spectrograms with $\mathcal{M}:y_i\rightarrow {y}'_i$\footnote{\href{https://librosa.org/doc/main/generated/librosa.feature.melspectrogram.html.}{Link: The librosa for audio to mel-spectrogram.}} as an intermediate means to facilitate the translation. The mel-scale results from the non-linear transformations of a Hz-scale, thereby emphasizing the human voice frequency (e.g., 40 to 1000 Hz) and suppressing the high-frequency instrument noise. 

\subsection{Asymmetric Fully-CNN Translator with Self Residual Attention}

In order to achieve the heterogeneous translation, we propose a novel self residual attention guided fully CNN translator with a pairwise disentanglement scheme, as shown in Fig. \ref{fig:illus}. We note that the generation of temporal sequences usually relies on RNNs, which can be challenging to train \cite{pascanu2013difficulty,xie2021deep}. In addition, connecting CNNs with RNNs using the fully connected layers is likely to lose the relative spatial and temporal information \cite{liu2020auto3d}. Considering the inherent discrepancy between $x_i$ and ${y}'_i$, we equip our $f$ with an asymmetric encoder-decoder structure. Since the input sequence of tagged-MRI $x_i$ usually has a fixed number of MR images, e.g., $T=26$, we use the 3D-CNN encoder \cite{liu2019dependency} for fast encoding. Notably, each 3D convolutional layer yields the same dimension in the temporal direction, where the dimension is reduced by half after each 3D MaxPooling operation to adaptively summarize the temporal information. For the decoder, we adopt a 2D-CNN with deconvolutional layers to synthesis $\tilde{y}'_i$. With the training pair $\{x_i, y_i\}$, the reconstruction loss can be a crucial supervision signal. We empirically adopt the mean square error (MSE) loss as:
\begin{align}
    \mathcal{L}_{rec} = ||\tilde{y}'_i-{y}'_i||_2^2 = ||\mathcal{F}(x_i)-\mathcal{M}(y_i)||_2^2.
\end{align}

Compared with conventional video analysis, tagged-MRI sequences contain highly redundant and irrelevant information for audio synthesis, since voice generation only involves a small region of interest, including the tongue and vocal cords. To specifically explore the moving muscular structures from tagged-MRI sequences, we explicitly emphasize the moving muscular structures related to speech production with an attention strategy. In particular, we denote the $t$-th time frame in each $x_i$ as $x_{i:t}$, $t\in\{1,\cdots,T\}$, and calculate the difference between two adjacent frames $d_{i:t}=|x_{i:t}-x_{i:t-1}|$ for $t\in\{2,\cdots,T\}$ as the residual frames to indicate the moving regions. Then, each residual frame $d_{i:t}\in\mathbb{R}^{H_x\times W_x}$ is processed with a self-trained attention network $\mathcal{A}$ to yield the fine-grained corresponding attention mask $a_{i:t}\in\mathbb{R}^{H_x\times W_x}$ for $t\in\{2,\cdots,T\}$. Then, the obtained attention mask $a_{i:t}$ is multiplied by the original tagged-MRI frame $x_{i:t}$ to generate the attentive frame $x'_{i:t}=a_{i:t}\otimes x_{i:t}$, which allows the network to filter the redundant static parts out. We adopt a conventional 2D encoder-decoder structure as $\mathcal{A}$, which is jointly optimized with $\mathcal{F}$, by minimizing $\mathcal{L}_{rec}$. Therefore, $\mathcal{A}$ is encouraged to adaptively learn to keep the essential information to guarantee synthesis performance, following a self-training fashion. Of note, we do not need an additional attention label to guide the training.

%propose to explicitly imposing a prior knowledge that the content part of speech tasks is similar from one subject to another. %We disentangle the word content and subject-specific factors by .  
 
\subsection{Pair-wise Disentanglement Training}

To exploit prior knowledge of the utterance content similarity in the latent space, an additional constraint is imposed in the latent space. We first disentangle the latent feature as an utterance-specific factor $u_i$ and a subject-specific factor $s_i$. Then, similar to deep metric learning \cite{liu2017adaptive}, the recorded $x_i$, speaking the same utterance, is encouraged to have similar latent representations as $u_i$. Empirically, although a multi-branch network can be used to split two parts, it could be inefficient for a fully-CNN framework. Instead, therefore, we propose to differentiate the specific channels in the feature representation from the tensor slice operation \cite{liu2020auto3d}\footnote{\href{https://pytorch.org/docs/stable/generated/torch.Tensor.select.html}{Link: Slicing the tensor in PyTorch.}}. 

For a pair of inputs $\{x_1,x_2\}$ with the same utterance, we explicitly enforce their $\{u_1,u_2\}$ to approximate each other. We opt to measure and minimize their discrepancy using the KL divergence with the reparameterization trick. In practice, we leverage the Gaussian prior and choose a few channels for the feature to denote the mean $\mu_i$ and variance $\sigma_i$, where $\mu_i$ has the same size as $\sigma_i$. We then make use of the reparametrization trick ${u_i}=\mu_i+\sigma_i\odot\epsilon_i$ to represent $u_i$ with $\mu_i$ and $\sigma_i$, where $\epsilon\in \mathcal{N}(0,I)$ \cite{liu2021dual}. The detailed KL divergence between ${u_1}$ and ${u_2}$ is given by:  
\begin{align} 
   \mathcal{L}_{KL}=-\frac{1}{2}\sum^{M}_{m=1}[1+{\rm log}\frac{\sigma_{1m}^2}{\sigma_{2m}^2} - \frac{\sigma_{1m}^2}{\sigma_{2m}^2} - \frac{(\mu_{1m} - \mu_{2m})^2}{\sigma_{2m}^2}], \label{eq:kl1}
\end{align}where $M$ represents the number of channels of the mean or variance ($M$=14 in our implementation). $\mathcal{L}_{KL}$ is only applied to the same utterance pairs.  

In parallel, $s_i$ is encouraged to inherit the subject-specific factors with an implicit complementary constraint \cite{liu2019feature,liu2021mutual}. By enforcing the information bottleneck, i.e., compact or low-dimensional latent feature \cite{liu2021mutual}, $s_i$ has to incorporate all the necessary complementary content (e.g., subject-specific style of the articulation) other than $u_i$ to achieve accurate reconstruction. Therefore, the generative task in the decoder is essentially taking $s_i$ conditioned on $u_i$, which models an utterance-conditioned spectrogram distribution, following a divide-and-conquer strategy \cite{che2019deep,liu2021domain}.

%, and thereby is easier than vanilla encoder-decoder modeling

\subsection{Adversarial Loss and Overall Training Protocol}

To further improve quality in our generated spectrograms, we leverage a GAN module. The discriminator $\mathcal{D}$ takes as input both the real mel spectrogram ${{y}'_i}=\mathcal{M}(y_i)$ the generated mel spectrogram $\tilde{{y}}'_i$, followed by identifying which is generated or real. The binary cross-entropy loss of the discriminator can be expressed as: 
\begin{align}
\mathcal{L}_{\mathcal{D}} = \mathbb{E}_{{y}'_i}\{\text{log}(\mathcal{D}({y}'_i))\} +  \mathbb{E}_{\tilde{{y}}'_i}\{\text{log}(1-\mathcal{D}(\tilde{{y}}'_i))\}.
\end{align}

In contrast, the translator attempts to confuse the discriminator, by yielding realistic spectrograms \cite{liu2021dual}. Notably, the translator does not involve real spectrograms in $\text{log}(\mathcal{D}({y}'_i))$ \cite{salimans2016improved}. Therefore, the translator can be trained, by optimizing the following objective: 
\begin{align}
\mathcal{L}_{\mathcal{F}:\mathcal{D}}= \mathbb{E}_{\tilde{{y}}'_i}\{-\text{log}(1-\mathcal{D}(\tilde{{y}}'_i))\}.
\end{align}  

In summary, we jointly optimize the following objectives for the translator, attention network, and discriminator: 
\begin{align}
    ^{\text{min}}_{~\mathcal{F}}~ \mathcal{L}_{rec} + \beta\mathcal{L}_{KL}+ \lambda\mathcal{L}_{\mathcal{F}:\mathcal{D}};~~~^{\text{min}}_{~\mathcal{A}}~  \mathcal{L}_{rec};~~~^{\text{min}}_{~\mathcal{D}}~  \mathcal{L}_{\mathcal{D}},
\end{align}where $\beta$ and $\lambda$ represent the weighting parameters. Of note, $\beta=0$ for the pairs with the different utterances.   

\begin{figure*}[t]
\begin{center} 
\includegraphics[width=1\linewidth]{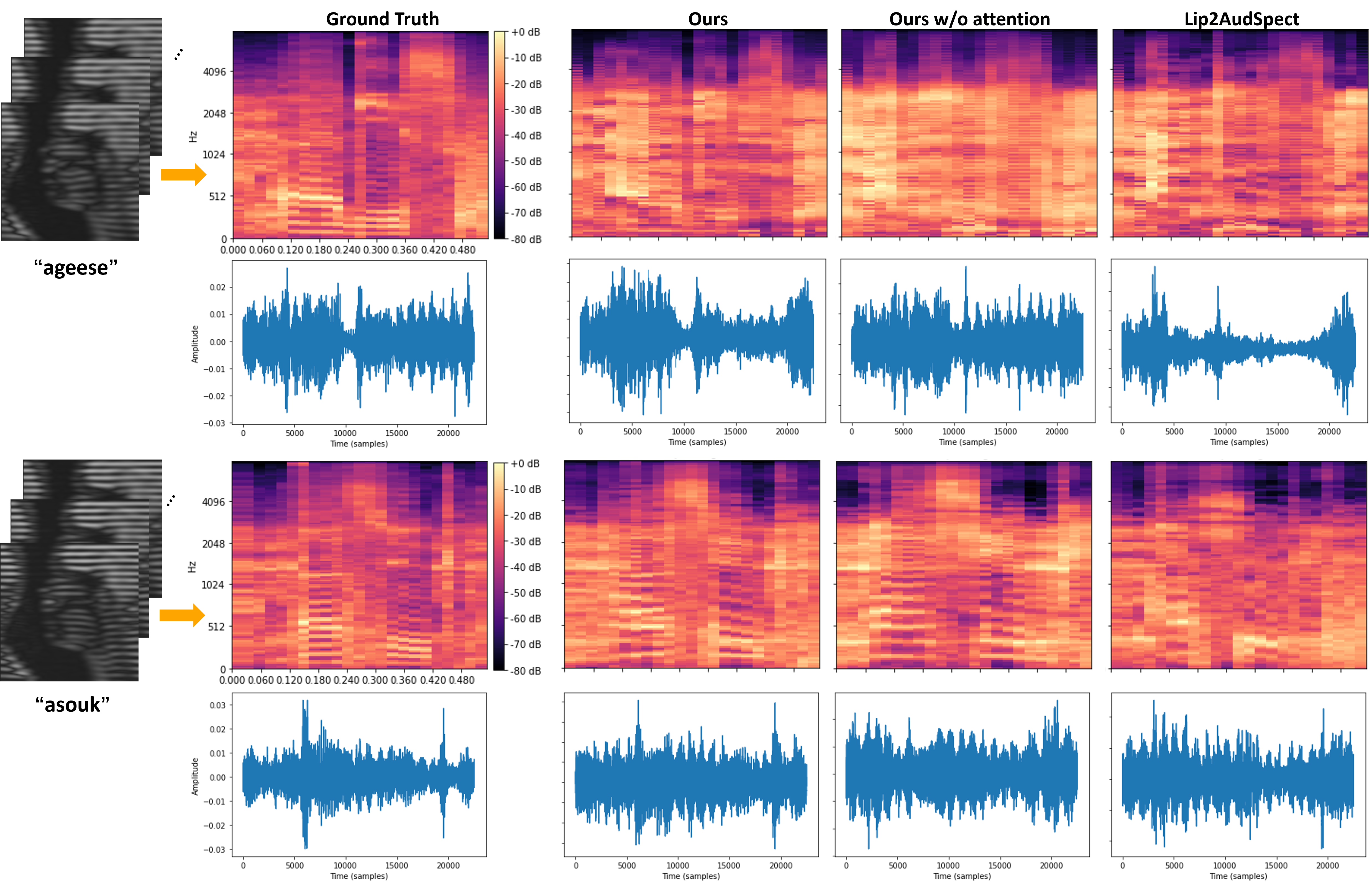}   
\end{center} 
\caption{Comparisons of our framework with Lip2AudSpect and ablation studies of the attention strategy. The audio files are attached in supplementary materials.}  
\label{fig:results}
\end{figure*}

In testing, the translator $\mathcal{F}$ is used to make inference, as shown in Fig.~\ref{fig:illus}; in addition, the pairwise inputs and the discriminator are not used. Thus, the pairwise framework and additional adversarial training do not affect the inference speed in implementation. Of note, it is straightforward to convert the spectrograms into waveforms, once we obtain the spectrograms\footnote{\href{https://librosa.org/doc/main/generated/librosa.feature.inverse.mel_to_audio.html}{Link: The librosa library for reversing mel-spectrogram to audio.}}.

\section{Experiments and Results}

To demonstrate the effectiveness of our framework, a total of 63 pairs of tagged-MRI sequences and audio waveforms were acquired, while a total of 43 participants performed a speech word ``asouk,” and a total of 20 participants performed a speech word ``ageese,” following a periodic metronome-like sound. A Siemens 3.0T TIM Trio system was used to acquire our data for which we used a 12-channel head coil and a 4-channel neck coil using a segmented gradient echo sequence~\cite{lee2013semi,xing20133d}. Imaging parameters are as follows: the field of view was 240mm$\times$240mm on the mid-sagittal slice with a resolution of 1.87mm$\times$1.87mm. Each tagged-MRI sequence had 26 time frames, which was then resized to 128$\times$128. The corresponding audio waveforms had their length varying from 21,832 to 24,175. To augment the datasets, we utilized a sliding window alongside the audio to crop a section with 21,000 time points, generating 100$\times$ audio data. Then, we used the Librosa library to convert all of the audio waveforms into mel spectrograms with the size of $64\times64$. In our evaluation, leave-one-out evaluation is used in a subject-independent manner. 
\begin{table}[t]
\centering 
\caption{Numerical comparisons in testing with leave-one-out evaluation. The best results are \textbf{bold}.} 
\resizebox{0.9\linewidth}{!}{
\begin{tabular}{l|c|c}
\hline
Methods& ~Corr2D for spectrogram $\uparrow$~ & ~PESQ for waveform $\uparrow$~\\\hline\hline
Lip2AudSpect \cite{akbari2018lip2audspec}        &{0.665}$\pm$0.014  &  1.235$\pm$0.021\\\hline

Ours                & \textbf{0.813}$\pm$0.010  &  \textbf{1.620}$\pm$0.017  \\\hline

Ours w/o Attention 	&0.781$\pm$0.012  &  1.517$\pm$0.025 \\
Ours w/o Pair-wise Disentangle 	& 0.798$\pm$0.010  &  1.545$\pm$0.019 \\
Ours w/o GAN      	&  {0.808}$\pm$0.012  &   1.586$\pm$0.023 \\\hline

\end{tabular}} 
\label{tabel:1} 
\end{table}

In practice, our encoder took five 3D convolutional layers, followed by the tensor slice and a decoder with four 2D deconvolutional layers. The rectified linear unit (ReLU) was used as an activation unit, while the sigmoid function was utilized to normalize the final output of each pixel. The attention network used a 2D encoder and decoder structure with four convolutional and four symmetric deconcolutional layers, which is followed by the $1\times1$ convolution with sigmoid unit. We used three convolutional layers and two fully connected layers with a sigmoid output as our discriminator. We separate the latent representation with 128 channels into three parts, i.e., 14 channels for both the mean and variance of the utterance-specific factors, and the remaining 100 channels for the subject-specific factors. Due to space limitations, we provide the detailed network structure in supplementary.

We implemented our framework using PyTorch and trained it on an NVIDIA V100 GPU, which took about 4.5 hours. In testing, the inference took only 0.5s. The learning rate was set at $lr_{\mathcal{F}}=10^{-3}$, $lr_{\mathcal{A}}=10^{-3}$, and $lr_{\mathcal{D}}=10^{-4}$ and the momentum was set at 0.5. To balance the involved optimization objectives, we used $\beta=0.5$ or 0 for the same or different utterance pairs, and set $\lambda=0.5$.

Fig.~\ref{fig:results} shows the qualitative results of our framework with and without the attention strategy and a comparison method. For comparison, we reimplemented Lip2AudSpect \cite{akbari2018lip2audspec} suitable to process tagged MRI data. We can see that our generated spectrogram and the recovered corresponding audio waveforms align better with the ground truth than the Lip2AudSpect, which uses a relatively sophisticated CNN-RNN-fully connected structure. We note that the 3D CNN used in Lip2AudSpect has a temporal window size of 5, which can only extract the local temporal correlation. Thus, Lip2AudSpect relies on RNN for long-term temporal modeling, which renders a difficulty in training on a limited number of datasets. To show the effectiveness of our self-attention strategy, we also provide the ablation study in Fig.~\ref{fig:results}, showing our superior performance over the comparison methods.

For quantitative evaluation, we followed \cite{akbari2018lip2audspec} to adopt 2D Pearson’s correlation coefficient (Corr2D) to measure the spectrogram synthesis quality in the frequency domain \cite{chi2005multiresolution}. In addition, the standard Perceptual Evaluation of Speech Quality (PESQ) was used to measure the quality of generated waveforms in the time domain \cite{recommendation2001perceptual}. The numerical comparisons among our framework, its ablation studies, and Lip2AudSpect \cite{akbari2018lip2audspec} are provided in Table \ref{tabel:1}. The standard deviation was reported by three random trials. Our framework outperformed Lip2AudSpect \cite{akbari2018lip2audspec} consistently. In addition, the synthesis performance was improved by the attention strategy, by a large margin. Furthermore, the performances of the pair-wise disentanglement and GAN loss were demonstrated as the ablation studies, showing their effectiveness in our overall network design to yield accurate synthesis results.

\begin{figure}[t!]
\begin{center} 
\includegraphics[width=1\linewidth]{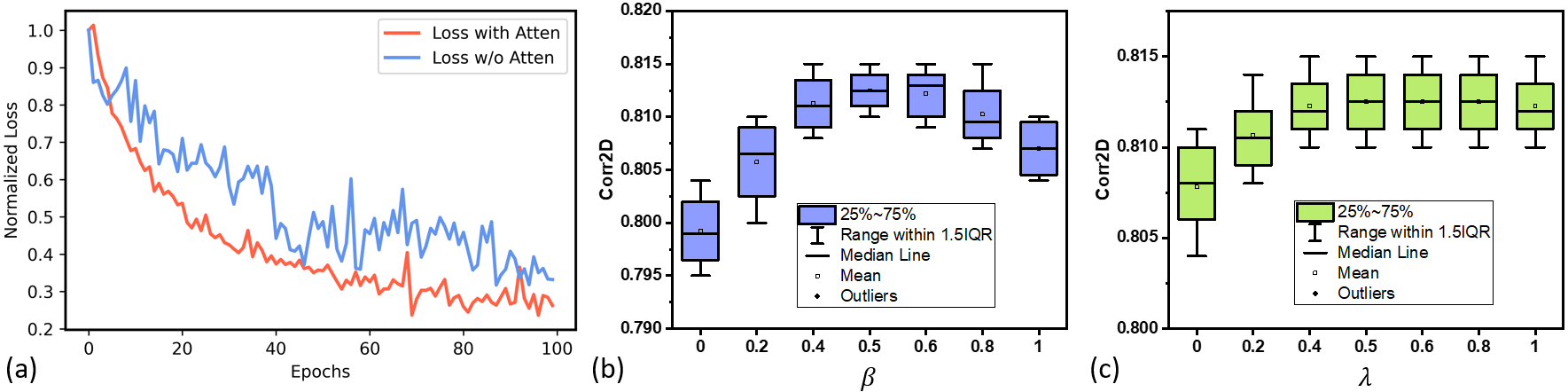}
\end{center}  
\caption{(a) Comparison of normalized loss using our framework with or without the self residual attention strategy. Sensitivity analysis of $\beta$ (b) and $\lambda$ (c).} 
\label{exp2}
\end{figure}

In Fig. \ref{exp2}(a), we show that our proposed self residual attention strategy achieves a stable loss decrease, via the information extraction module, without being distracted by the redundant surrounding areas. As shown in Figs. \ref{exp2}(b) and \ref{exp2}(c), the performance is relatively stable for $\beta\in[0.4,0.6]$ to impose pair-wise disentanglement. In addition, the GAN loss was effective for $\lambda\in[0.4,0.8]$.    

\section{Discussion and Conclusion}

In this work, we proposed a novel framework to synthesize spectrograms from tagged-MRI sequences. The audio waveforms can also be obtained from the synthesized spectrograms. In particular, we proposed an efficient fully convolutional asymmetry translator with help of a self residual attention strategy to specifically focus on the moving muscular structures for speech production.~Additionally, we used a pairwise correlation of the samples with the same utterances with a latent space representation disentanglement scheme.~Furthermore, we incorporated an adversarial training approach with GAN to yield improved results on our generated spectrograms.~Our experimental results showed that our framework was able to successfully synthesize spectrograms (and clear waveforms) from tagged-MRI sequences, outperforming the Lipreading based method. Therefore, our framework offered the potential to help clinicians improve treatment strategies for patients with speech-related disorders. In future work, we will investigate the use of full three-dimensional plus time tagged-MRI sequences as well as tracking information from tagged-MRI to achieve spectrogram synthesis.

\section*{Acknowledgements}
 
This work is supported by NIH R01DC014717, R01DC018511, and R01CA133015.

\bibliographystyle{splncs04}
\bibliography{egbib}

\begin{thebibliography}{10}
\providecommand{\url}[1]{\texttt{#1}}
\providecommand{\urlprefix}{URL }
\providecommand{\doi}[1]{https://doi.org/#1}

\bibitem{akbari2018lip2audspec}
Akbari, H., Arora, H., Cao, L., Mesgarani, N.: Lip2{audspec}: Speech
  reconstruction from silent lip movements video. In: ICASSP. pp. 2516--2520.
  IEEE (2018)

\bibitem{che2019deep}
Che, T., Liu, X., Li, S., Ge, Y., Zhang, R., Xiong, C., Bengio, Y.: Deep
  verifier networks: Verification of deep discriminative models with deep
  generative models. AAAI  (2021)

\bibitem{chi2005multiresolution}
Chi, T., Ru, P., Shamma, S.A.: Multiresolution spectrotemporal analysis of
  complex sounds. The Journal of the Acoustical Society of America
  \textbf{118}(2),  887--906 (2005)

\bibitem{chung2016lip}
Chung, J.S., Zisserman, A.: Lip reading in the wild. In: ACCV. pp. 87--103.
  Springer (2016)

\bibitem{ephrat2017vid2speech}
Ephrat, A., Peleg, S.: Vid2speech: speech reconstruction from silent video. In:
  ICASSP. pp. 5095--5099. IEEE (2017)

\bibitem{goodfellow2017deep}
Goodfellow, I., Bengio, Y., Courville, A.: Deep learning (adaptive computation
  and machine learning series). MIT Press  (2017)

\bibitem{griffin1984signal}
Griffin, D., Lim, J.: Signal estimation from modified short-time fourier
  transform. IEEE Transactions on acoustics, speech, and signal processing
  \textbf{32}(2),  236--243 (1984)

\bibitem{he2020classification}
He, G., Liu, X., Fan, F., You, J.: Classification-aware semi-supervised domain
  adaptation. In: Proceedings of the IEEE/CVF Conference on Computer Vision and
  Pattern Recognition Workshops. pp. 964--965 (2020)

\bibitem{he2020image2audio}
He, G., Liu, X., Fan, F., You, J.: Image2audio: Facilitating semi-supervised
  audio emotion recognition with facial expression image. In: Proceedings of
  the IEEE/CVF Conference on Computer Vision and Pattern Recognition Workshops.
  pp. 912--913 (2020)

\bibitem{lee2013semi}
Lee, J., Woo, J., Xing, F., Murano, E.Z., Stone, M., Prince, J.L.:
  Semi-automatic segmentation of the tongue for 3{D} motion analysis with
  dynamic {MRI}. In: ISBI. pp. 1465--1468. IEEE (2013)

\bibitem{liu2021mutual}
Liu, X., Chao, Y., You, J.J., Kuo, C.C.J., Vijayakumar, B.: Mutual information
  regularized feature-level frankenstein for discriminative recognition. IEEE
  TPAMI  (2021)

\bibitem{liu2020auto3d}
Liu, X., Che, T., Lu, Y., Yang, C., Li, S., You, J.: Auto3d: Novel view
  synthesis through unsupervisely learned variational viewpoint and global {3d}
  representation. In: ECCV. pp. 52--71. Springer (2020)

\bibitem{liu2019dependency}
Liu, X., Guo, Z., You, J., Kumar, B.V.: Dependency-aware attention control for
  image set-based face recognition. IEEE Transactions on Information Forensics
  and Security  \textbf{15},  1501--1512 (2019)

\bibitem{liu2021domain}
Liu, X., Hu, B., Jin, L., Han, X., Xing, F., Ouyang, J., Lu, J., Fakhri, G.E.,
  Woo, J.: Domain generalization under conditional and label shifts via
  variational bayesian inference. IJCAI  (2021)

\bibitem{liu2019feature}
Liu, X., Li, S., Kong, L., Xie, W., Jia, P., You, J., Kumar, B.: Feature-level
  frankenstein: Eliminating variations for discriminative recognition. In:
  CVPR. pp. 637--646 (2019)

\bibitem{liu2017adaptive}
Liu, X., Vijaya~Kumar, B., You, J., Jia, P.: Adaptive deep metric learning for
  identity-aware facial expression recognition. In: CVPR. pp. 20--29 (2017)

\bibitem{liu2021dual}
Liu, X., Xing, F., Prince, J.L., Carass, A., Stone, M., El~Fakhri, G., Woo, J.:
  Dual-cycle constrained bijective vae-gan for tagged-to-cine magnetic
  resonance image synthesis. In: ISBI. pp. 1448--1452. IEEE (2021)

\bibitem{liu2022structure}
Liu, X., Xing, F., Prince, J.L., Stone, M., El~Fakhri, G., Woo, J.:
  Structure-aware unsupervised tagged-to-cine mri synthesis with self
  disentanglement. In: Medical Imaging 2022: Image Processing. vol. 12032, pp.
  470--476. SPIE (2022)

\bibitem{liu2022cmri2spec}
Liu, X., Xing, F., Stone, M., Prince, J.L., Kim, J., El~Fakhri, G., Woo, J.:
  Cmri2spec: Cine mri sequence to spectrogram synthesis via a pairwise
  heterogeneous translator. In: ICASSP 2022-2022 IEEE International Conference
  on Acoustics, Speech and Signal Processing (ICASSP). pp. 1481--1485. IEEE
  (2022)

\bibitem{liu2022tagged}
Liu, X., Xing, F., Stone, M., Prince, J.L., Kim, J., El~Fakhri, G., Woo, J.:
  Tagged-mri to audio synthesis with a pairwise heterogeneous deep translator.
  The Journal of the Acoustical Society of America  \textbf{151}(4),
  A133--A133 (2022)

\bibitem{liu2021generative}
Liu, X., Xing, F., Stone, M., Zhuo, J., Reese, T., Prince, J.L., El~Fakhri, G.,
  Woo, J.: Generative self-training for cross-domain unsupervised
  tagged-to-cine mri synthesis. In: International Conference on Medical Image
  Computing and Computer-Assisted Intervention. pp. 138--148. Springer (2021)

\bibitem{liu2020symmetric}
Liu, X., Xing, F., Yang, C., Kuo, C.C.J., Fakhri, G.E., Woo, J.:
  Symmetric-constrained irregular structure inpainting for brain mri
  registration with tumor pathology. In: International MICCAI Brainlesion
  Workshop. pp. 80--91. Springer (2020)

\bibitem{michelsanti2021overview}
Michelsanti, D., Tan, Z.H., Zhang, S.X., Xu, Y., Yu, M., Yu, D., Jensen, J.: An
  overview of deep-learning-based audio-visual speech enhancement and
  separation. IEEE/ACM Transactions on Audio, Speech, and Language Processing
  (2021)

\bibitem{pascanu2013difficulty}
Pascanu, R., Mikolov, T., Bengio, Y.: On the difficulty of training recurrent
  neural networks. In: ICML. pp. 1310--1318. PMLR (2013)

\bibitem{recommendation2001perceptual}
Recommendation, I.T.: Perceptual evaluation of speech quality {PESQ}): An
  objective method for end-to-end speech quality assessment of narrow-band
  telephone networks and speech codecs. Rec. ITU-T P. 862  (2001)

\bibitem{salimans2016improved}
Salimans, T., Goodfellow, I., Zaremba, W., Cheung, V., Radford, A., Chen, X.:
  Improved techniques for training gans. NIPS  \textbf{29},  2234--2242 (2016)

\bibitem{wang2021automated}
Wang, J., Liu, X., Wang, F., Zheng, L., Gao, F., Zhang, H., Zhang, X., Xie, W.,
  Wang, B.: Automated interpretation of congenital heart disease from
  multi-view echocardiograms. Medical Image Analysis  \textbf{69},  101942
  (2021)

\bibitem{xie2021deep}
Xie, W., Liang, L., Lu, Y., Luo, H., Liu, X.: Deep {3D-CNN} for depression
  diagnosis with facial video recording of self-rating depression scale
  questionnaire. JBHI  (2021)

\bibitem{xing2022brain}
Xing, F., Liu, X., Kuo, J., Fakhri, G., Woo, J.: Brain mr atlas construction
  using symmetric deep neural inpainting. IEEE Journal of Biomedical and Health
  Informatics  (2022)

\bibitem{xing20133d}
Xing, F., Woo, J., Murano, E.Z., Lee, J., Stone, M., Prince, J.L.: {3D} tongue
  motion from tagged and cine {MR} images. In: MICCAI. pp. 41--48. Springer
  (2013)

\bibitem{yu2021reconstructing}
Yu, Y., Shandiz, A.H., T{\'o}th, L.: Reconstructing speech from real-time
  articulatory mri using neural vocoders. In: 2021 29th European Signal
  Processing Conference (EUSIPCO). pp. 945--949. IEEE (2021)

\end{thebibliography}

\end{document}